\documentclass[aps,prb,showpacs,tightenlines,longbibliography,twocolumn,nofootinbib,nobibnotes,superscriptaddress]{revtex4-1}
\usepackage{amsmath,amssymb,amsfonts,bm}
\usepackage{graphicx}
\usepackage{epstopdf}
\usepackage{dcolumn}
\usepackage{mathrsfs}
\usepackage{tgtermes}
\usepackage[colorlinks=true,linkcolor=blue,citecolor=blue, urlcolor=blue,bookmarks=false]{hyperref}
\bibpunct{[}{]}{,}{n}{}{}

\begin{document}
\title{Edge states in a non-Hermitian topological crystalline insulator}
\date{\today }
\author{Qiu-Yue Xu}
\affiliation{Department of Physics, Hubei University, Wuhan 430062, China}
\author{Feng Liu}
\affiliation{Department of Physics, Ningbo University, Ningbo 315211, China}
\author{Chui-Zhen Chen}
\affiliation{Institute for Advanced Study and School of Physical Science and Technology, Soochow University, Suzhou 215006, China}
\author{Dong-Hui Xu}\email{donghuixu@hubu.edu.cn}
\affiliation{Department of Physics, Hubei University, Wuhan 430062, China}

\begin{abstract}
Breaking Hermiticity in topological systems gives rise to intriguing phenomena, such as the exceptional topology and the non-Hermitian skin effect. In this work, we study a non-Hermitian topological crystalline insulator sitting on the Kekul\'{e} texture-modulated honeycomb lattice with balanced gain and loss. We find that the gaplessness of the topological edge states in the non-Hermitian system is insensitive to edge geometries under moderate strength of gain and loss, unlike the cases of Hermitian topological crystalline insulators that depend on edge geometries crucially. We focus on two types of gain and loss configurations, which are $PT$-symmetric and $PT$-asymmetric, respectively.
For the $PT$-symmetric configuration, the Dirac point of the topological edge states in the Hermitian molecular-zigzag-terminated ribbons splits into a pair of exceptional points. The edge gap in the Hermitian armchair-terminated ribbons vanishes and a Dirac point forms as far as moderate gain and loss is induced. The band gaps of edge and bulk states in the Hermitian armchair-terminated ribbons close simultaneously for the $PT$-asymmetric configuration. 
\end{abstract}

\maketitle

\section{Introduction}

Following the discovery of topological insulators~\cite{RevModPhys.82.3045,RevModPhys.83.1057}, the search for novel symmetry-protected topological phases of quantum matter has become one of the central themes in condensed matter physics. Topological crystalline insulators (TCIs)~\cite{PhysRevLett.106.106802} are such a novel topological phase protected by crystalline symmetries. The gaplessness of Dirac surface states in TCIs usually depends on geometries of surface terminations crucially, which makes them more fragile than those in conventional topological insulators protected by time-reversal symmetry. Recently, the pursuit of topological phases has been extended to non-Hermitian systems~\cite{PhysRevB.84.205128,alvarez2018topological,PhysRevX.8.031079,AdvPhys2020,PhysRevX.9.041015}. Non-Hermitian systems have unique topological properties beyond the Hermitian framework owing to complex-valued energy spectra. One of the most salient characteristics of non-Hermitian systems is the emergence of exceptional points~\cite{moiseyev2011non,berry2004physics, Heiss_2012}, where pairs of eigenvalues and the corresponding eigenvectors coalesce. The exceptional point introduces several fascinating topological phenomena~\cite{RevModPhys.93.015005}, such as the exceptional rings~\cite{zhen2015spawning,PhysRevLett.120.146402,PhysRevLett.118.045701,cerjan2019experimental,PhysRevB.99.121101,taoliu2021}, the bulk Fermi arcs and half-integer topological charges~\cite{kozii2017non,Zhou1009,PhysRevB.98.035141,PhysRevLett.125.227204}. 

\begin{figure}[t]
	\includegraphics[width=8cm]{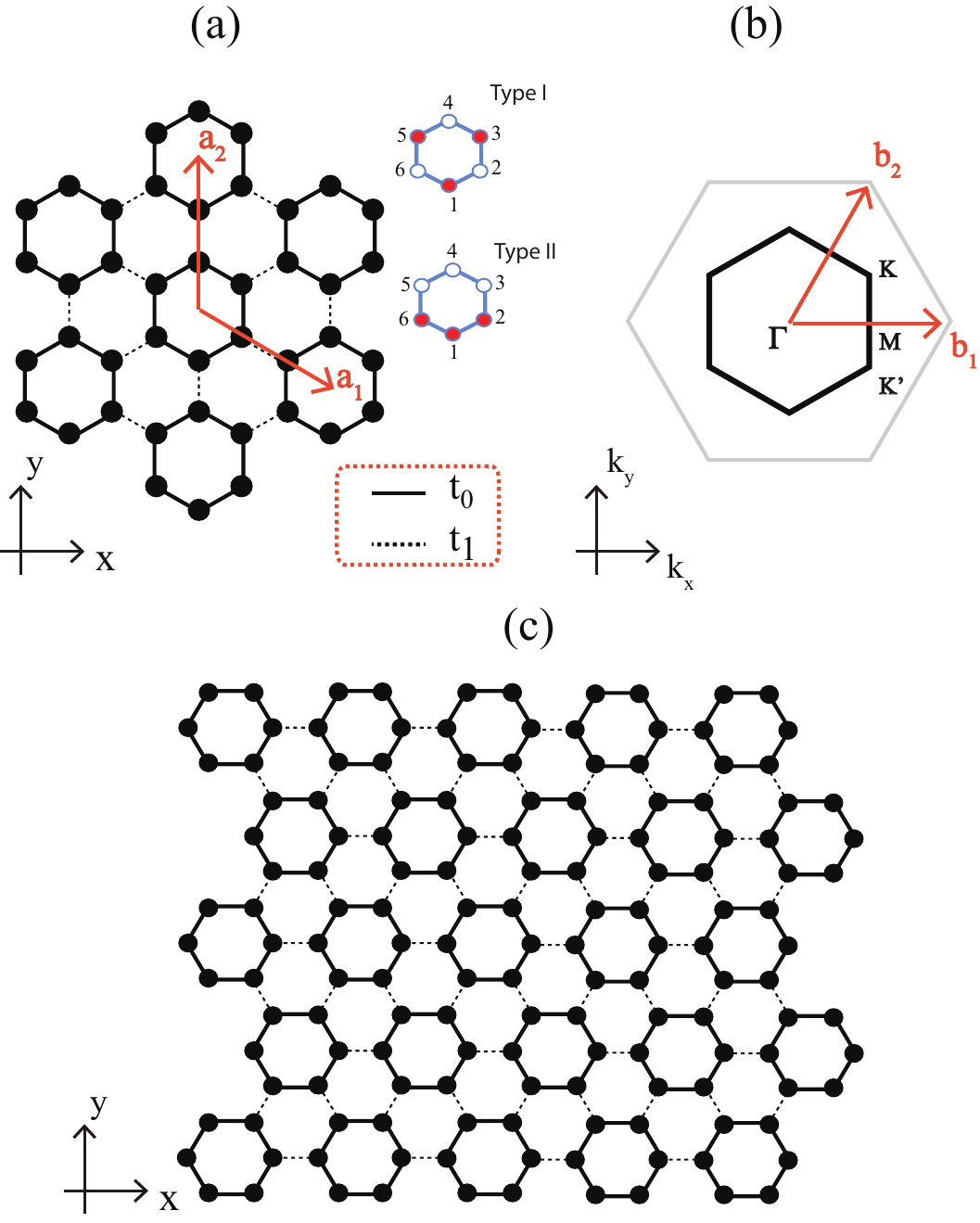} \caption{(a) Schematic of the non-Hermitian honeycomb lattice with a Kekul\'{e} bond texture. The thick solid (thin dashed) black lines represent the intracellular bonds (intercellular bonds). The sublattices in the unit cell are indexed as $1,2,...,6$, accordingly. The inset shows the two types of gain and loss configurations, which are the $PT$-symmetric type I and $PT$-asymmetric type II. The red filled and the unfilled circles in the hexagons denote denote the lattices with gain $i\gamma$ and the lattices with loss $-i\gamma$, respectively.  The orange arrows show the enlarged lattice vectors $\vec{a}_{1}$ and $\vec{a}_{2}$. (b) Reduced Brillouin zone for the Kekul\'{e} lattice. The gray hexagon represents the original Brillouin zone for the honeycomb lattice. (c) A Kekul\'{e} lattice showing two type of boundaries compatible with unit cell defined in (a). The armchair boundary is along the $x$-direction and molecular-zigzag boundary is along the $y$-direction.}
	\label{fig1}
\end{figure}

Interestingly, the bulk-edge correspondence, one of the essential concepts in topological materials, has been proved to be subtle in non-Hermitian systems~\cite{Xiong_2018,PhysRevLett.121.026808,PhysRevLett.121.086803,PhysRevB.99.155431,PhysRevB.99.201103,PhysRevB.99.081103,PhysRevA.99.052118,PhysRevB.100.054105,PhysRevLett.124.056802,PhysRevB.100.161105,PhysRevLett.123.066404,PhysRevB.101.195147}.  
This is partially due to the appearance of abnormal geometric structures (points/rings/disks), eigenstates and eigenvalues can coalesce in the complex energy space. For instance, the point gap topology of complex spectra can lead to the unique skin effect in non-Hermitian systems~\cite{PhysRevLett.77.570,PhysRevLett.116.133903,PhysRevLett.121.086803,PhysRevLett.121.026808,PhysRevLett.121.136802,PhysRevLett.124.086801,PhysRevLett.125.126402,PhysRevLett.125.186802,PhysRevLett.125.226402,PhysRevLett.124.250402,PhysRevResearch.2.022062}. Although many unique non-Hermitian properties of topological systems have been revealed in the context, the non-Hermitian properties of TCIs of various geometries, especially for edge terminations, are still worthy of exploration.  
As a prototype system, we use the two-dimensional (2D) honeycomb lattice decorated with the modulated Kekul\'{e} hopping texture, as displayed in Fig.~\ref{fig1}(a). Due to the asymmetric hopping texture, each unit cell in the Kekul\'{e} lattice consists of six sites, in contrast to the ideal honeycomb lattice. The hopping texture in the Kekul\'{e} lattice couples the valley degrees of freedom and gaps out the Dirac cones~\cite{PhysRevB.62.2806,PhysRevLett.98.186809,PhysRevB.80.233409,CHEIANOV20091499,Gamayun_2018}. The Kekul\'{e} texture can be realized experimentally for various solid-state materials, such as in the molecular graphene \cite{gomes2012designer} and Lithium-intercalated graphene~\cite{PhysRevLett.126.206804}. Remarkably, the Kekul\'{e}-texture-modulated honeycomb lattice has been recognized as a 2D TCI when the intercellular hopping is greater than the intracellular hopping~\cite{kariyado2017topological,PhysRevLett.122.086804,PSJ.86.123707}, and the gapless topological edge states are protected by mirror symmetry $M_y$ and chiral symmetry. Very recently, the topological edge states are observed in the artificial Kekul\'{e} lattice by positioning the CO molecules on Cu(111) surface~\cite{PhysRevLett.124.236404}. Moreover, the TCI on the Kekul\'{e} lattice exhibits higher-order topology~\cite{noh2018topological,PhysRevLett.122.086804,PhysRevLett.123.053902,JPSJ.88.104703,PhysRevB.101.241109} and the corner states have been detected in photonic systems~\cite{noh2018topological}, electrical circuits~\cite{PhysRevLett.123.053902}, and acoustic systems~\cite{PhysRevLett.125.255502}. The Kekul\'{e} lattice has attracted intensive research interest, and the previous studies focus only on the Hermitian case. The study on the non-Hermitian Kekul\'{e} lattice is still lacking.

 In this work, we study the Kekul\'{e} lattice subject to balanced gain and loss, i.e., the gain and the loss have a same amplitude. In the Hermitian case, the topological edge states of the TCI on the Kekul\'{e} lattice are sensitive to edge geometries. In particular, for the molecular-zigzag terminated Kekul\'{e} lattice that preserves mirror symmetry $M_y$, the edge states are gapless, and the Dirac point of the edge states is pinned at zero energy thanks to chiral symmetry. In the armchair-terminated Kekul\'{e} lattice, the edge states are gapped because of the mirror symmetry breaking. We consider two types of balanced gain and loss configurations that introduce non-Hermiticity, i.e., the $PT$ symmetric type I and the $PT$ asymmetric type II as displayed in the inset of Fig.~\ref{fig1}(a). For both configurations, the bulk gap is reduced by increasing the strength of gain and loss. For the $PT$ symmetric configuration, the edge states of molecular-zigzag-terminated ribbon show a pair of exceptional points and have a finite imaginary energy. In the armchair-terminated ribbon, the energy gap of the edge states closes, and a Dirac point forms while tuning the strength of $PT$ symmetric gain and loss. Furthermore, the non-Hermiticity-induced Dirac point in the armchair-terminated ribbon splits into a pair of exceptional points as further increasing the strength of gain and loss. For the $PT$ asymmetric gain and loss, the energy spectra of the bulk and edge states become complex once turning on the gain and loss. The edge and bulk gaps close simultaneously for the $PT$ asymmetric configuration.

 This paper is organized as follows: In Sec.~\ref{The Model}, we introduce the tight-binding model of the Kekul\'{e} hopping texture modulated honeycomb lattice in the presence of balanced gain and loss. Then, we present the non-Hermitian effects on the TCI phase of the Kekul\'{e} lattice in Sec.~\ref{nonhermitian} for both $PT$-symmetric and $PT$-asymmetric gain and loss configurations. Finally, a brief summary is presented in Sec.~\ref{Conclusion}.

\section{Model}
\label{The Model}

\begin{figure*}[t]
	\includegraphics[width=14cm]{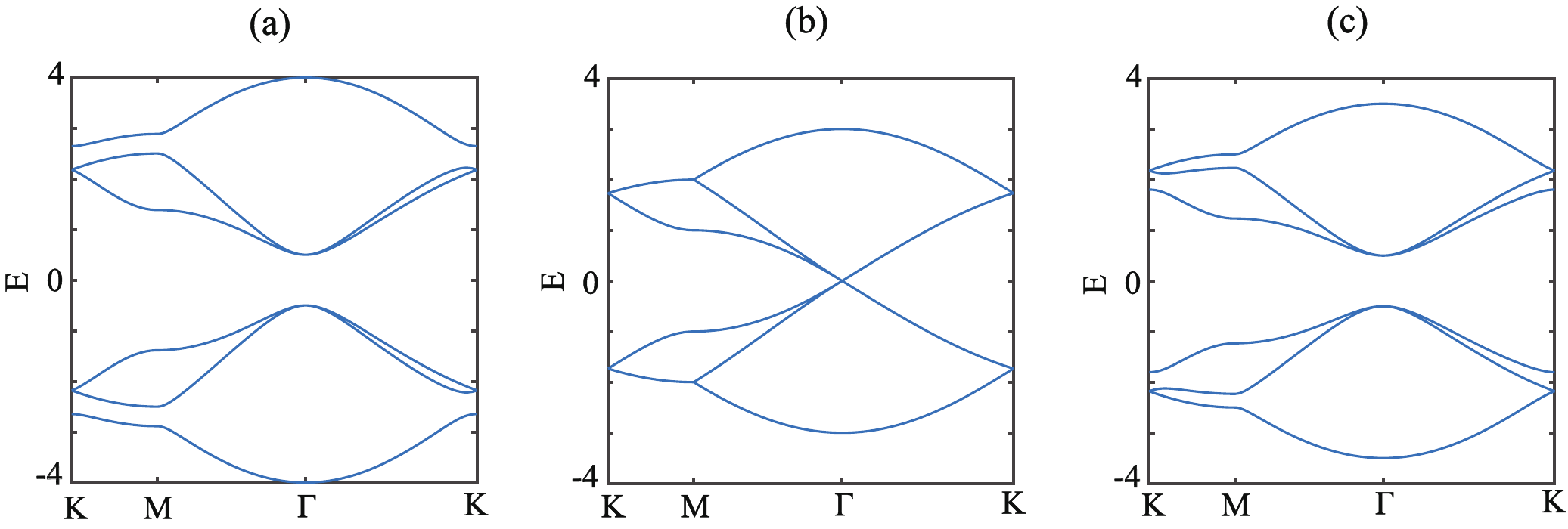} \caption{ Bulk energy band structure of the tight-binding model on the Kekul\'{e} hopping texture modulated honeycomb lattice with hopping parameters $t_0$ and $t_1$. (a) $(t_{0},t_{1})=(1.5,1)$. The topologically trivial phase. (b) $(t_{0},t_{1})=(1,1)$. The critical semimetal phase. (c) $(t_{0},t_{1})=(1,1.5)$. The topologically nontrivial TCI phase.}%
	\label{fig2}
\end{figure*}

\begin{figure}[t]
	\includegraphics[width=8.0cm]{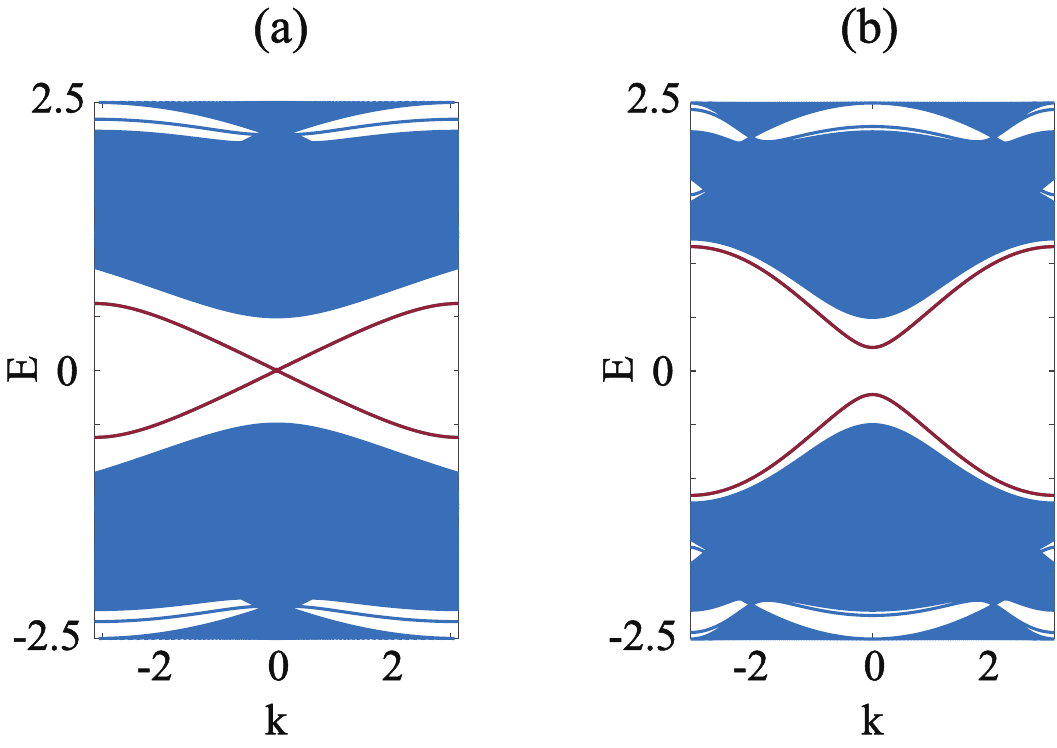} \caption{ Boundary geometry dependent edge states in the 2D TCI phase. (a) The energy spectrum for a molecular-zigzag terminated ribbon. The gapless edge modes (red lines) appear inside the bulk energy gap. (b) The spectrum for an armchair-terminated ribbon. The edge modes (red lines) inside the bulk gap display an energy gap. The hopping parameters are $(t_{0},t_{1})=(1,1.5)$ in both (a) and (b).}%
	\label{fig3}
\end{figure}

\begin{figure*}[t]
	\includegraphics[width=14cm]{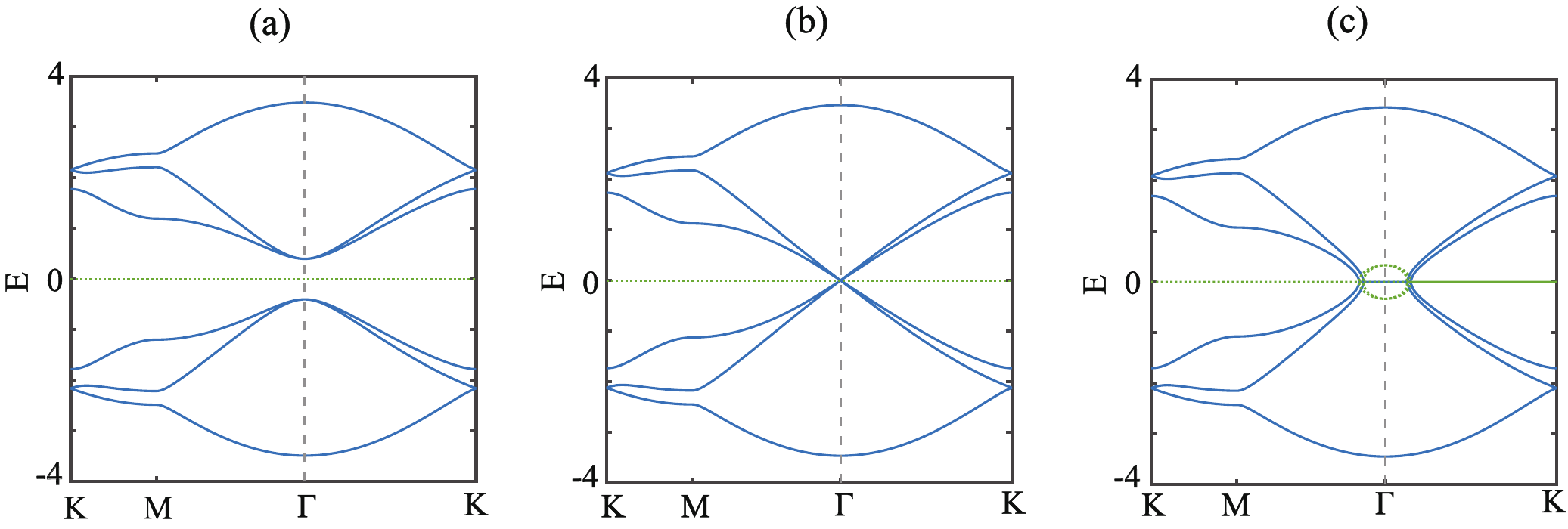} \caption{ Bulk band structure of the non-Hermitian TCI phase in the present of first configuration of gain and loss. (a) $\gamma=0.3$, (b) $\gamma=0.5$, (c) $\gamma=0.6$. The blue lines correspond to the real part of the energy, and the dashed green lines are the imaginary part. The spectrum becomes complex and the flat bands appear when $\gamma>0.5$. The $PT$ symmetric gain and loss per unit cell is given by $( {i\gamma, -i\gamma, i\gamma, -i\gamma, i\gamma, -i\gamma})$. The hopping parameters are $(t_{0},t_{1})=(1,1.5)$.}%
	\label{fig4}
\end{figure*}

\begin{figure*}[hptb]
	\includegraphics[width=17cm]{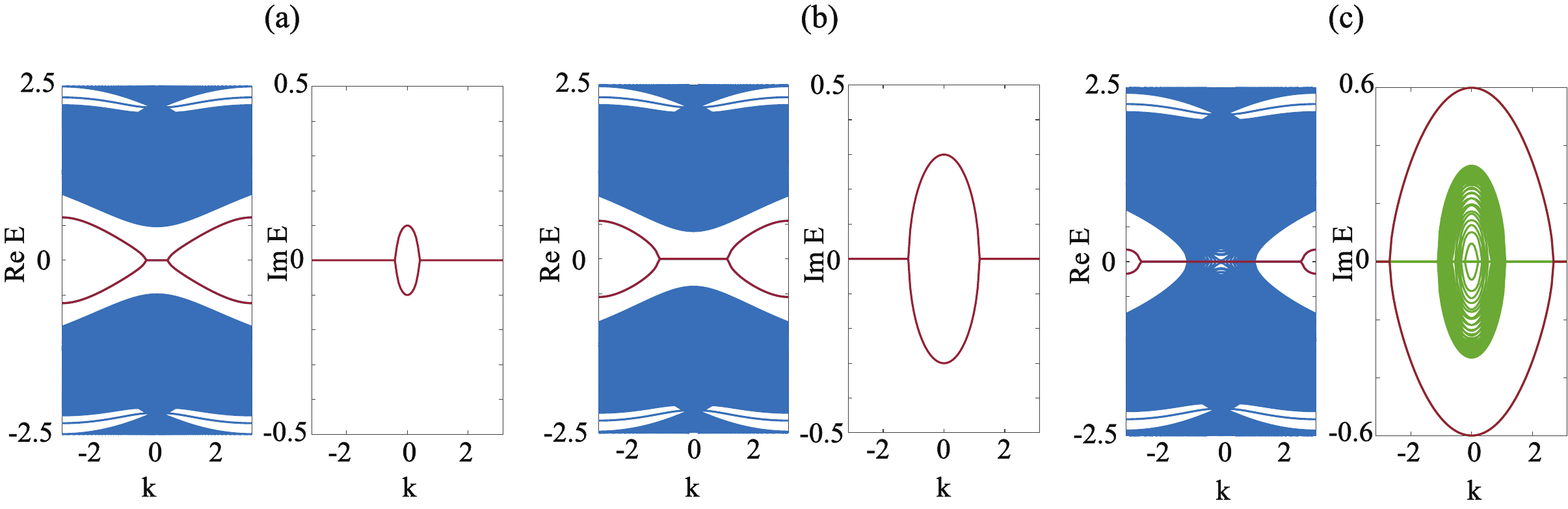} \caption{The energy spectra for a molecular-zigzag terminated ribbon in the non-Hermitian TCI phase, and the $PT$ symmetric gain and loss in the unit cell are given by $ ({i\gamma, -i\gamma, i\gamma, -i\gamma, i\gamma, -i\gamma})$. (a) $\gamma=0.1$, (b) $\gamma=0.3$ and (c) $\gamma=0.6$. The red lines mark the real and imaginary part of energy spectrum of edge states. The green curves mark the imaginary energy of the bulk states. The hopping parameters are $(t_{0},t_{1})=(1,1.5)$.}%
	\label{fig5}
\end{figure*}

\begin{figure*}[t]
	\includegraphics[width=17cm]{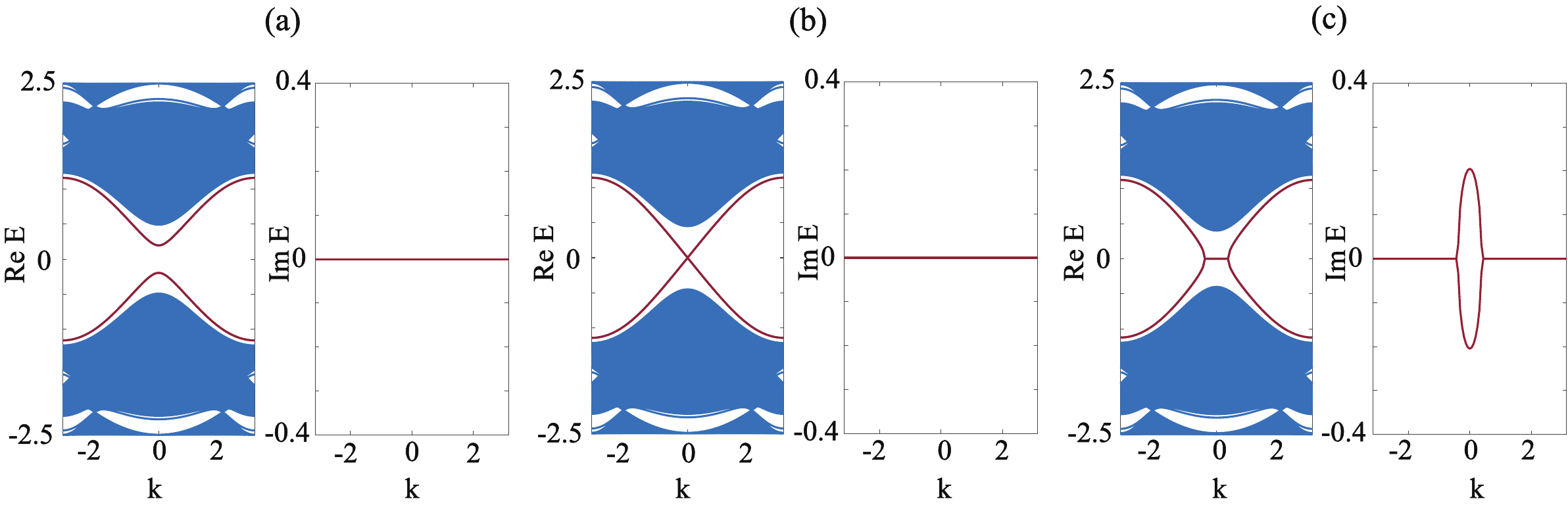} \caption{The spectra for an armchair terminated ribbon in the non-Hermitian TCI phase, and the $PT$ symmetric gain and loss in each unit cell is given by $ ({i\gamma, -i\gamma, i\gamma, -i\gamma, i\gamma, -i\gamma})$. The right parts of (a)-(c) show the imaginary parts of the edge spectra. (a) $\gamma=0.1$. (b) $\gamma=0.2192$. This is the critical points at which the edge gap is closed. (c) $\gamma=0.3$. The red lines mark the real and imaginary part of energy spectrum of edge states. The hopping parameters are $(t_{0},t_{1})=(1,1.5)$.}%
	\label{fig6}
\end{figure*}

\begin{figure}[t]
	\includegraphics[width=8cm]{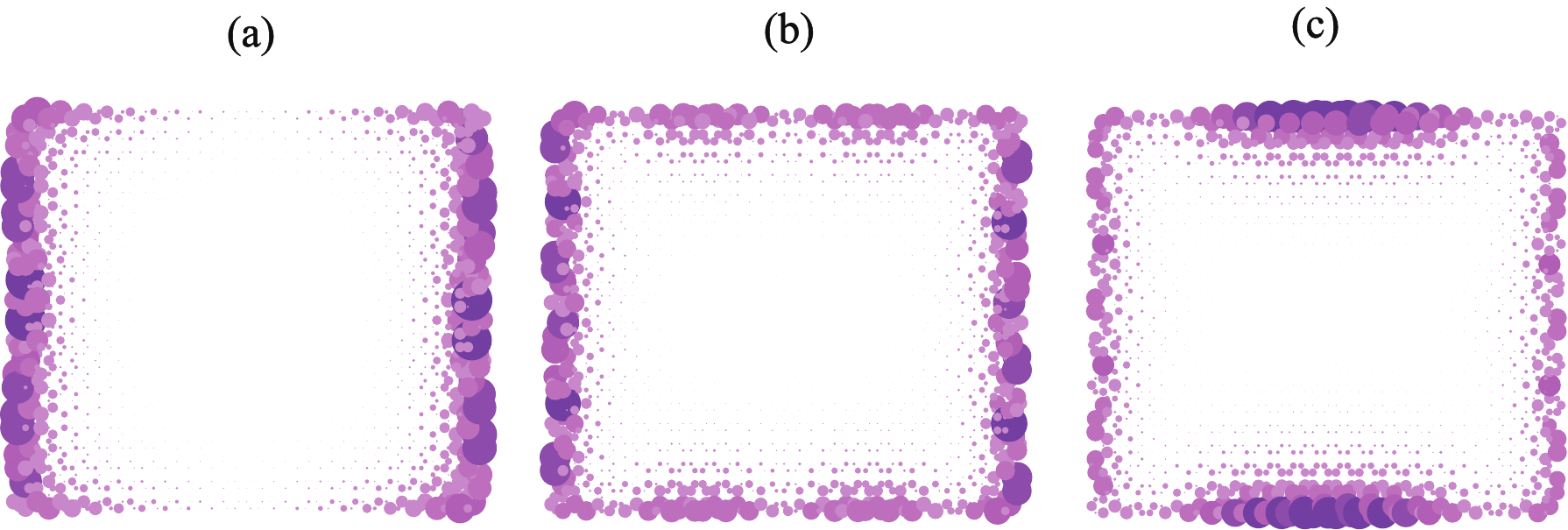} \caption{The sample comprised of $20 \times 20$ unit cells supporting the edge states. The molecular-zigzag boundary is along the $y$-direction, and the armchair boundary is along the $x$ direction. (a) The distribution of edge states with energy around $0.15$ in the Hermitian case. (b) The distribution of edge states with energy around $0.28$, which is above the edge gap but still in the bulk gap. (c) In the non-Hermitian case of $\gamma=0.2$, the distribution of edge states with energy around $0.15$, which is located in the original edge gap.}%
	\label{fig7}
\end{figure}

 The Kekul\'{e} lattice can be viewed as a honeycomb lattice with an alternating bond texture, as depicted in Fig. \ref{fig1}(a). Owing to the alternating bond texture, the unit cell is enlarged, and there are six sublattices in a unit cell. The two primitive lattice vectors are defined as $\mathbf{a}_1=a(3/2,-\sqrt{3}/2)$ and $\mathbf{a}_2=a(0,\sqrt{3})$ with $a$ the lattice constant. Correspondingly, the two unit vectors in the reciprocal lattice are $\mathbf{b}_1=\frac{2\pi}{3a}(2,0)$ and $\mathbf{b}_2=\frac{2\pi}{3a}(1,\sqrt{3})$. We introduce two types of nearest-neighbor hopping parameters consist of the intracellular hopping $t_{0}$ and the intercellular hopping $t_{1}$~[See the orange dashed rectangle in Fig.~\ref{fig1}(a)]. Then the Hermitian tight-binding Hamiltonian reads
\begin{equation}\label{H0real}
H_0=-\sum_{\langle i,j\rangle}t _{i,j}c_{i}^{\dag}c_{j},
\end{equation}
where $\langle i,j \rangle$ represents the nearest-neighbors pairs on the honeycomb lattice, $c_{i}^{\dag}$ and $c_{i}$ are the creation and annihilation operators at the site $i$. The hopping parameter $t _{i,j}=t_{0}>0$ if $i$ and $j$ are connected by a solid bond and belong to the same cell in Fig.~\ref{fig1}(a), and $t _{i,j}=t_{1}>0$ if $i$ and $j$ are connected by a dashed bond and belong to the adjacent cells. On this basis of $\big( c_{\mathbf{k},1},c_{\mathbf{k},2},c_{\mathbf{k},3},c_{\mathbf{k},4},c_{\mathbf{k},5},c_{\mathbf{k},6}\big)^T$, the Hamiltonian matrix in momentum space can be written as
 \begin{widetext}
 	
 	\begin{equation}
 		H_0(\mathbf{k})=-\left(\!
 		\begin{array}{cccccc}
 			\!0  & t_{0} & 0 & t_{1}e^{i\mathbf{k} \cdot \mathbf{a}_{2}} & 0 & t_{0}\! \\
 			\!t_{0} & 0  & t_{0} & 0 & t_{1}e^{-i\mathbf{k} \cdot \mathbf{a}_{1}} & 0\! \\
 			\!0 & t_{0} & 0  & t_{0} & 0 & t_{1}e^{-i \mathbf{k} \cdot (\mathbf{a}_{1}+\mathbf{a}_{2})}\! \\
 			\!t_{1}e^{-i\mathbf{k} \cdot \mathbf{a}_{2}} & 0 & t_{0} & 0 & t_{0} & 0\! \\
 			\!0  & t_{1}e^{i\mathbf{k} \cdot \mathbf{a}_{1}} & 0 & t_{0} & 0 & t_{0}\!  \\
 			\!t_{0} & 0 &  t_{1}e^{i\mathbf{k} \cdot (\mathbf{a}_{1}+\mathbf{a}_{2})}  & 0 & t_{0} & 0\! \\
 		\end{array}%
 		\!\right).
 		\label{H0}
 	\end{equation}
 \end{widetext}
 
 Before introducing the non-Hermitian effect, it is useful to discuss the symmetries and band structures of the Hermitian Kekul\'{e} lattice. Equations \ref{H0real} and \ref{H0} preserve time-reversal symmetry and chiral symmetry, which are two internal symmetries. For this spinless system, time-reversal symmetry is expressed as $\mathcal{T}=\mathcal{K}$ with $\mathcal{K}$ the complex conjugate. Whereas chiral symmetry $\mathcal{C}$ is defined as
 \begin{equation}\label{CS}
 	\mathcal{C}H_0(\mathbf{k})\mathcal{C}^{-1}=-H_0(\mathbf{k}), \;\mathcal{C}^2=1.
 \end{equation} 
 where $\mathcal{C}=\sigma_z\oplus(\sigma_0\otimes\sigma_z)$ on the basis of Eq.~(2), $\sigma_{z}$ is the $z$-component of the Pauli matrix vector and $\sigma_0$ denotes the two by two identity matrix. Besides the two internal symmetries, the Hamiltonian respects inversion symmetry that can be described by the form of matrix as follows
	\begin{equation}
	\mathcal{P}=\left(\!
	\begin{array}{cccccc}
     \!0 & 0 & 0 & 1 & 0 & 0\! \\
	\!0 & 0  & 0 & 0 & 1 & 0\! \\
		\!0 &0 & 0& 0 & 0 &1\! \\
		\!1 & 0 & 0 & 0 & 0 & 0\! \\
 		\!0  & 1 & 0 & 0 & 0 & 0\!  \\
 		\!0 & 0 &  1  & 0 & 0 & 0\! \\
\end{array}%
\!\right).\nonumber
 \end{equation}
In addition, the Hamiltonian is also invariant under the crystalline symmetries that include a six-fold rotation symmetry $C_6$ as well as the two inequivalent mirror reflection symmetries $M_x$ and $M_y$. 
 
 The competition between intercellular and intracellular hoppings plays an essential role in controlling the gap-opening of the bulk energy bands that determines the system's topology. So, in some sense, the Kekul\'{e} lattice can be viewed as a 2D extension of the Su-Schrieffer-Heeger~(SSH) model~\cite{SSHmodel}. When $t_0\neq t_1$, an energy gap is opened at $\Gamma$ point, and the system is an insulator at half-filling. At $t_0=t_1$, the system reduces to the ideal honeycomb lattice hosting a 2D Dirac semimetal. The system is a normal insulator for $t_0/t_1>1$.  As the ratio $t_0/t_1$ decreases, a topological phase transition occurs at the critical point $t_0/t_1=1$, and the system becomes a 2D TCI for $t_0/t_1<1$~\cite{kariyado2017topological,PhysRevLett.122.086804,PSJ.86.123707}. 
 Figure~\ref{fig2} displays the bulk energy spectra for various ratios of $t_0/t_1$. In Fig.~\ref{fig2}(a), we show the energy spectrum of the topologically trivial insulator when $t_0>t_1$. The energy gap at $\Gamma$ point closes when $t_0/t_1=1$, as shown in Fig.~\ref{fig2}(b). Whereas Fig.~\ref{fig2}(c) shows the spectrum of the 2D TCI. For demonstrating the topological edge states of the TCI, we plot the energy spectra of the Kekul\'{e} nanoribbons with the molecular-zigzag boundary and the armchair boundary in Figs.~\ref{fig3}(a) and \ref{fig3}(b), respectively. The topological edge states are sensitive to the edge geometries of the sample. We can see that the molecular-zigzag terminated Kekul\'{e} lattice supports the gapless edge states, and the Dirac point formed by the band crossing of edge states is protected by both the mirror reflection symmetry $M_y$ and the chiral symmetry $\mathcal{C}$. $M_y$ is broken in the Kekul\'{e} lattice with the armchair boundary, and the edge states are gapped therefore. Note that the boundary-dependent edge states have been observed in a recent experiment~\cite{PhysRevLett.124.236404}.

 \section{Non-Hermitian effects}
 \label{nonhermitian}
 For studying non-Hermitian effects on the Kekul\'{e} lattice, we consider that each unit cell suffers balanced gain and loss, which is described by the following Hamiltonian
 \begin{equation}
 	\Delta H=i\gamma\sum_{i,\alpha,\beta}\left(c_{i,\alpha}^{\dag}c_{i,\alpha}- c_{i,\beta}^{\dag}c_{i,\beta}\right),
 	\label{H}
 \end{equation}
 where $\gamma$ denotes the gain and loss strength, and $\alpha$ and $\beta$ label the sublattices suffering gain ($i\gamma$) and loss ($-i\gamma$), respectively. Here we focus on two distinct configurations of gain and loss, which are marked as type I and type II in the inset of Fig.~\ref{fig1}(a). For type I, $\alpha=1,3,5$ and $\beta=2,4,6$, while for type II, $\alpha=1,2,6$ and $\beta=3,4,5$. The total non-Hermitian Hamiltonian is given as $H=H_0+\Delta H$. The non-Hermiticity induced by gain and loss has led to intriguing topological phase transitions and phenomena in 1D and 2D SSH models~\cite{PhysRevA.89.062102,PhysRevLett.116.133903,PhysRevA.96.032103,PhysRevA.95.053626,pan2018photonic,PhysRevB.98.094307,PhysRevLett.121.213902,PhysRevA.97.042118,PhysRevB.99.155431,PhysRevLett.123.165701,PhysRevA.100.032102,PhysRevA.99.012113}.
 
 In the non-Hermitian case, chiral symmetry transforms the Hamiltonian in the following way~\cite{PhysRevX.9.041015}
\begin{equation}
	\mathcal{C}H^\dagger(\mathbf{k})\mathcal{C}^{-1}=-H(\mathbf{k}).
\end{equation} 
This equation is consistent with Eq.~(\ref{CS}) for Hermitian systems where $H^\dagger=H$. The two types of gain and loss configurations preserve chiral symmetry, which ensures symmetric energy spectra of the non-Hermitian TCI. 

\subsection{Type I configuration  of gain and loss}
\label{Pattern 1}

\begin{figure*}[t]
	\includegraphics[width=14cm]{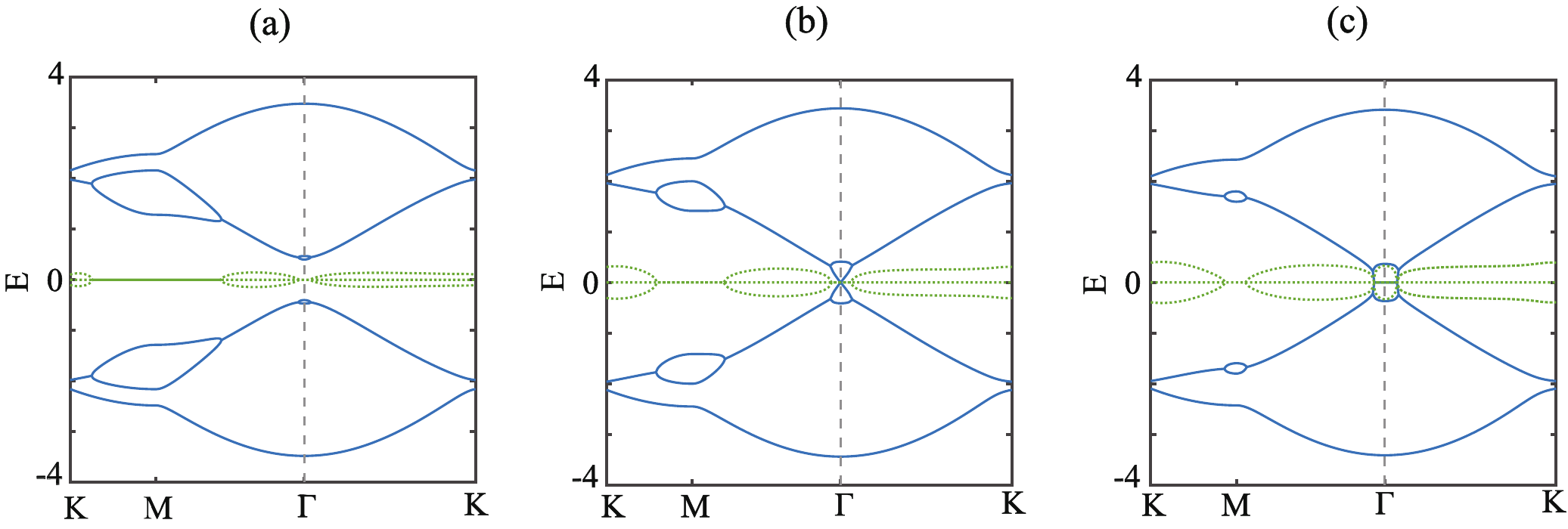} \caption{Bulk energy band structure for the non-Hermitian 2D TCI phase with  $(t_{0},t_{1})=(1,1.5)$. The gain and loss in each unit cell  that breaks $PT$ symmetry is given by $ ({i\gamma, i\gamma, -i\gamma, -i\gamma, -i\gamma, i\gamma})$. (a) $\gamma=0.3$, (b) $\gamma=0.5$, (c) $\gamma=0.6$. The blue lines are the real part of the energy, and the dashed green lines describe the imaginary part. }%
	\label{fig8}
\end{figure*}

\begin{figure*}[t]
	\includegraphics[width=17cm]{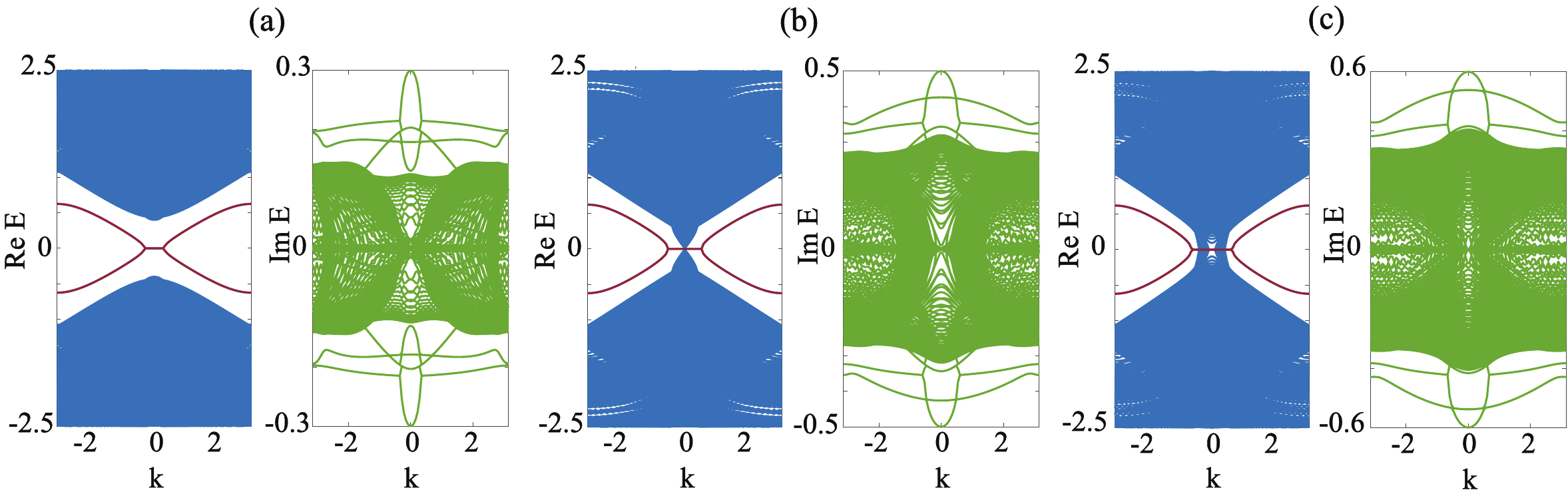} \caption{The energy spectra for a molecular-zigzag terminated ribbon in the presence of $PT$ asymmetric gain and loss in the unit cell, which is given by $ ({i\gamma, i\gamma, -i\gamma, -i\gamma, -i\gamma, i\gamma})$. (a) $\gamma=0.3$, (b) $\gamma=0.5$ and (c) $\gamma=0.6$. The red lines mark the real part of energy spectrum of edge states, while the green lines represent the imaginary parts of the bulk and edge states. The hopping parameters are $(t_{0},t_{1})=(1,1.5)$.}%
	\label{fig9}
\end{figure*}

\begin{figure*}[t]
	\includegraphics[width=17cm]{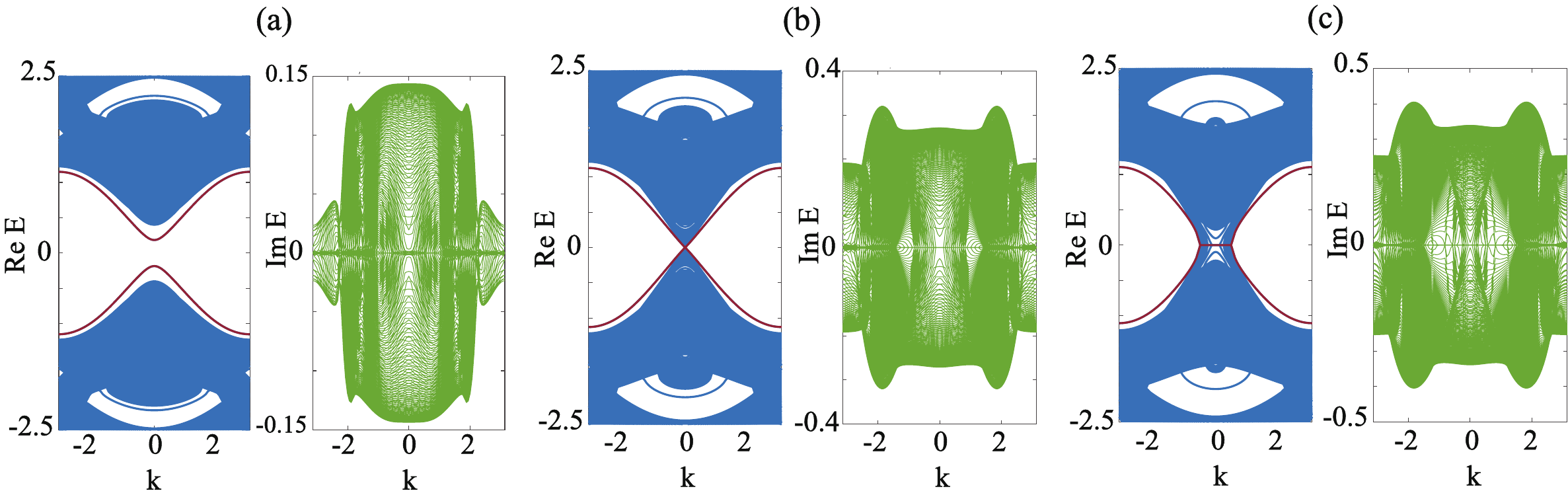} \caption{The energy spectra for an armchair terminated ribbon in the presence of $PT$ asymmetric gain and loss in the unit cell, which is given by $ ({i\gamma, i\gamma, -i\gamma, -i\gamma, -i\gamma, i\gamma})$. (a) $\gamma=0.3$, (b) $\gamma=0.5$ and (c) $\gamma=0.6$. The red lines mark the real part of energy spectrum of edge states, while the green lines represent the imaginary parts of the bulk and edge states. The hopping parameters are $(t_{0},t_{1})=(1,1.5)$.}%
	\label{fig10}
\end{figure*}

In this subsection, we consider type I configuration of gain and loss that preserves $PT$-symmetry, which can be described as $\Delta H=i\gamma \sigma_z\oplus(\sigma_0\otimes\sigma_z)$. In the presence of type I configuration of gain and loss, time-reversal symmetry is broken. However, the combination of inversion symmetry and time-reversal symmetry $\mathcal{P}\otimes \mathcal{T}$ is preserved. Therefore, the bulk energy spectrum remains real unless the $PT$-symmetry is spontaneously broken~\cite{bender1998real}. In the following of this article, we focus on the regime of $t_{0}<t_{1}$ where the system is in the TCI phase with topological gap opening at the $\Gamma$ point. In particular, the non-Hermitian Hamiltonian $H(\mathbf{k})$ has purely imaginary eigenvalues around the $\Gamma$ point when $\gamma>|t_{1}-t_{0}|$.
In Fig.~\ref{fig4}, we plot the bulk band structures for different values of $\gamma$. We can see that a topological phase transition occurs when tuning $\gamma$. In specific, the band gap gradually decreases to zero with increasing $\gamma$, and bulk Dirac cones are formed at the $\Gamma$ point for the critical point value $\gamma=|t_{1}-t_{0}|$. When $\gamma$ is further increased, the energy spectrum develops rings of exceptional points around the $\Gamma$ point, within which the real part of spectrum exhibits flat bands pinned at zero energy, while the imaginary part has a finite value, as displayed in Fig.~\ref{fig4}(c). Our results indicate that, in addition to nodal Hermitian systems~\cite{PhysRevB.104.L201104}, gapped Hermitian systems can also spawn exceptional points by introducing $PT$ symmetric anti-Hermitian terms.

Besides the non-Hermitian effect on the bulk band structure, the topological edge states of the 2D TCI phase also show interesting phenomena. As shown in Figs.~\ref{fig5}(a) and \ref{fig5}(b), the gapless edge states in the molecular-zigzag terminated TCI are robust against the balanced gain and loss if $\gamma<|t_{1}-t_{0}|$. However, the Dirac point of the edge states splits into two exceptional points connected by flat bands. Meanwhile, the eigenvalues of edge states have finite imaginary parts although the bulk spectrum is real. The imaginary part of edge spectrum increases as $\gamma$ increases, as displayed in Fig.~\ref{fig5}(b). When $\gamma>|t_1-t_0|$, the edge states are mixed with the bulk states as the bulk gap closes. The energy spectrum of the bulk states also has an imaginary part, whose magnitude is smaller than that of edge states, as depicted in Fig.~\ref{fig5}(c) .

The non-Hermitian effect on the edge states of the armchair-terminated TCI is even more striking. In the absence of gain and loss, the edge states are gapped owing to the mirror symmetry $M_y$ breaking. Turning on the gain and loss, the edge gap decreases with the increase of $\gamma$ but both the edge and the bulk energy spectra keep real, as shown in Fig.~\ref{fig6}(a). The edge gap is eventually closed and a non-Hermiticity induced Dirac point is formed when $\gamma$ is large enough, as depicted in Fig.~\ref{fig6}(b). As $\gamma$ is further increased, the formed Dirac point splits into two separated exceptional points [ see Fig.~\ref{fig6}(c) ]. In addition, as shown in Fig.~\ref{fig6}(c), the real part of edge spectrum shows a flat band and the imaginary part is finite within the range of the flat band. 

The effect of gain and loss on the edge states can be understood by a two-band effective edge model Hamiltonian, which reads $H_\text{eff}=-v_\text{F}k\tau_z+\Delta\tau_x+i\gamma\tau_y$, where $v_\text{F}$ denotes the Fermi velocity of the edge sates, $\tau_{x,y,z}$ are the Pauli matrices acting on the subspace formed by the edge states, and $\Delta$ represents the band gap of the edge states. The eigenvalues of this Hamiltonian are $E_\pm=\pm\sqrt{v^2_\text{F}k^2+\Delta^2-\gamma^2}$. Apparently, the edge gap at $\Gamma$ point determined by $\Delta$ will be closed when $\gamma=\Delta$. The eigenvalues $E_\pm$ become purely imaginary for $\gamma^2>(v^2_\text{F}k^2+\Delta^2)$. The flat bands pinned at zero energy appear since the real part of $E_\pm$ is zero in the regime determined by $\gamma^2>(v^2_\text{F}k^2+\Delta^2)$.

In order to further demonstrate the non-Hermitian effect on the edge states, we consider a finite square-shaped Kekul\'{e} lattice sample under the open boundary conditions along both the $x$ and $y$ directions. This finite size sample has two types of edges, the molecular-zigzag edge along the $y$ direction and the armchair edge along the $x$ direction. In the Hermitian case, when the chemical potential is located in the edge gap of the armchair boundaries, the molecular-zigzag edges show a finite probability density of electrons, as displayed in Fig.~\ref{fig7}(a). When the chemical potential is shifted out of the edge gap of the armchair-terminated ribbon but still within the bulk gap, the probability density of electrons is finite for both molecular-zigzag and armchair edges [see Fig.~\ref{fig7}(b)]. In the non-Hermitian case, even for a chemical potential located in the edge gap, by increasing $\gamma$ the probability distribution pattern will change from that shown in Fig.~\ref{fig7}(a) to the pattern exhibited in Fig.~\ref{fig7}(c), which suggests that the distribution of edge states can be controlled by tuning $\gamma$.

\subsection{Type II configuration of gain and loss}
\label{Pattern 2}
For comparison, we consider type II configuration of gain and loss that is $PT$ asymmetric [see the inset of Fig.~\ref{fig1}(a)], where the sublattices marked by 3, 4, 5 have the imaginary potential $-i\gamma$ and the sublattices 1, 2, 6 have $i\gamma$. The non-Hermitian Hamiltonian is no longer $PT$-invariant in presence of the type II configuration, thus the bulk energy spectrum becomes complex once turning on $\gamma$. Figure~\ref{fig8} plots the bulk spectrum for several different values of $\gamma$. We can see that the imaginary part of spectrum increases while the size of real bulk gap reduces to zero as $\gamma$ increases. We also plot the energy spectrum of a molecular-zigzag terminated sample for different values of $\gamma$ in Fig.~\ref{fig9}. The Dirac point of gapless edge states again splits into a pair of exceptional points. For the armchair boundary, type II configuration of gain and loss also reduces the edge gap as type I configuration does. Meanwhile, as shown in Fig.~\ref{fig10}, the edge gap and bulk gap close simultaneously as $\gamma$ raises, which is in contrast to the $PT$-symmetric gain and loss. In that case, the edge gap is closed before the bulk gap closes. It is suggested that we can use the $PT$ symmetric gain and loss to close the edge gap of the 2D TCI phase in experiments, meanwhile keep the bulk spectrum real and gapped, for the Kekul\'{e} lattice with the armchair boundary.

\section{Conclusion}
\label{Conclusion}
 To summarize, we have studied the 2D TCI phase in the honeycomb lattice with Kekul\'{e}-like hopping texture under balanced gain and loss.
 Particularly, we consider two types of gain and loss configurations that are $PT$ symmetric and $PT$ asymmetric, respectively. 
 We found both types of gain and loss configurations can close the bulk gap. However, the bulk spectrum remains real and gapped before the spontaneous $PT$-symmetry breaking occurs for the $PT$ symmetric gain and loss. In contrast, the bulk spectrum becomes complex once we introduce the $PT$ asymmetric gain and loss configurations. The edge states are dramatically affected by the two types of gain and loss configurations. The $PT$ symmetric gain and loss drives the Dirac point of edge states in the molecular-zigzag-terminated sample to split into a pair of exceptional points. The edge gap in the armchair terminated sample can be closed by the $PT$ symmetric gain and loss and a Dirac point forms. As the gain and loss strength further increases, the non-Hermiticity induced Dirac point also splits into two separated exceptional points before the bulk gap is closed. The $PT$ asymmetric gain and loss can also drive the Dirac point to split into exceptional points for the molecular-zigzag boundary. In the case of the armchair boundary, the edge gap and the bulk gap are simultaneously closed by the $PT$ asymmetric gain and loss. In a word, the TCI becomes more robust and nonsensitive to the edge geometries in the presence of moderate $PT$ symmetric gain and loss. With rapid progress in the experimental implementation of non-Hermiticity in artificial systems, we believe these exotic non-Hermitian phenomena uncovered in the 2D TCI state will be soon demonstrated in future experiments. The exceptional points in the edge spectrum might find their application in the design of $PT$-symmetric topological insulator lasers~\cite{harari2018topological,bandres2018topological} with only lasing edge modes.

\section*{Acknowledgments}
The authors acknowledge the support by the NSFC (under Grant Nos. 12074108, and 11974256), the NSF
of Jiangsu Province (under Grant No. BK20190813) and the Priority Academic Program Development (PAPD) of Jiangsu
Higher Education Institution. F. Liu acknowledges the financial support by the Research Starting Funding of Ningbo
University, NSFC Grant No. 12074205, and NSFZP Grant No. LQ21A040004.

\bibliography{bibfile}
\end{document}